\documentclass[11pt]{article}
\usepackage{arxiv}	
\usepackage{url}      
\usepackage{graphicx,times}
\usepackage{caption}
\usepackage{subcaption}
\usepackage{color}
\usepackage{natbib}
\usepackage{graphicx}
\usepackage{color}
\usepackage{hyperref}

\title{Satin Non-Woven Fabrics for Designing of Self-Regulating Breathable Building Skins}

\author{Saied Zarrinmehr \\
	Computer Science and Engineering Department,\\ Texas A\&M University, College Station, TX, 77831\\
	\texttt{szarinmehr@gmail.com } \\
 \And
 \href{https://orcid.org/0000-0003-3618-4166}{\includegraphics[scale=0.06]{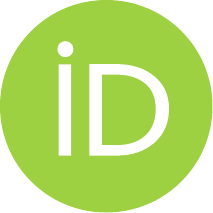}\hspace{1mm}Ergun Akleman}\thanks{Joint with Computer Science and Engineering Department.} \\
	Visual Computing \& Computational Media,\\ Texas A\&M University, College Station, TX, 77831\\
	\texttt{ergun@tamu.edu} \\
	\And
 Tahir Cagin \\
	Material Science and Engineering,\\ Texas A\&M University, College Station, TX, 77831\\
	\texttt{tcagin@tamu.edu} \\
	}

\renewcommand{\shorttitle}{Satin Non-Woven Breatable Fabrics}

\begin{document}

\maketitle

\thispagestyle{empty}

\begin{abstract}
In this paper, we introduce the concept of 2-way 2-fold genus-1 non-woven fabrics that can be used to design self-regulating breathable building skins. The advantage of non-woven structures over woven structures for breathable skin design is that they can completely be closed to stop air exchange. We have developed a theoretical framework for such non-woven structures starting from the mathematical theory of biaxial 2-fold  Genus-1 woven fabrics. By re-purposing a mathematical notation that is used to describe 2-fold  2-way 2-fold genus-1 woven fabrics, we identify and classify non-woven fabrics. Within this classification, we have identified a special subset that corresponds to satin woven fabrics and allows for maximum air exchange. Any other subset of non-woven structures that correspond to other classical 2-way 2-fold genus-1 fabrics, such as plain or twill, will allow for less air exchange. We also show that there exists another subset of satin non-woven fabrics that can provide the biggest openings.
\end{abstract}

\begin{figure*}[htpb]
    \centering
    \begin{subfigure}[t]{0.32\textwidth}
        \includegraphics[width=1.0\textwidth]{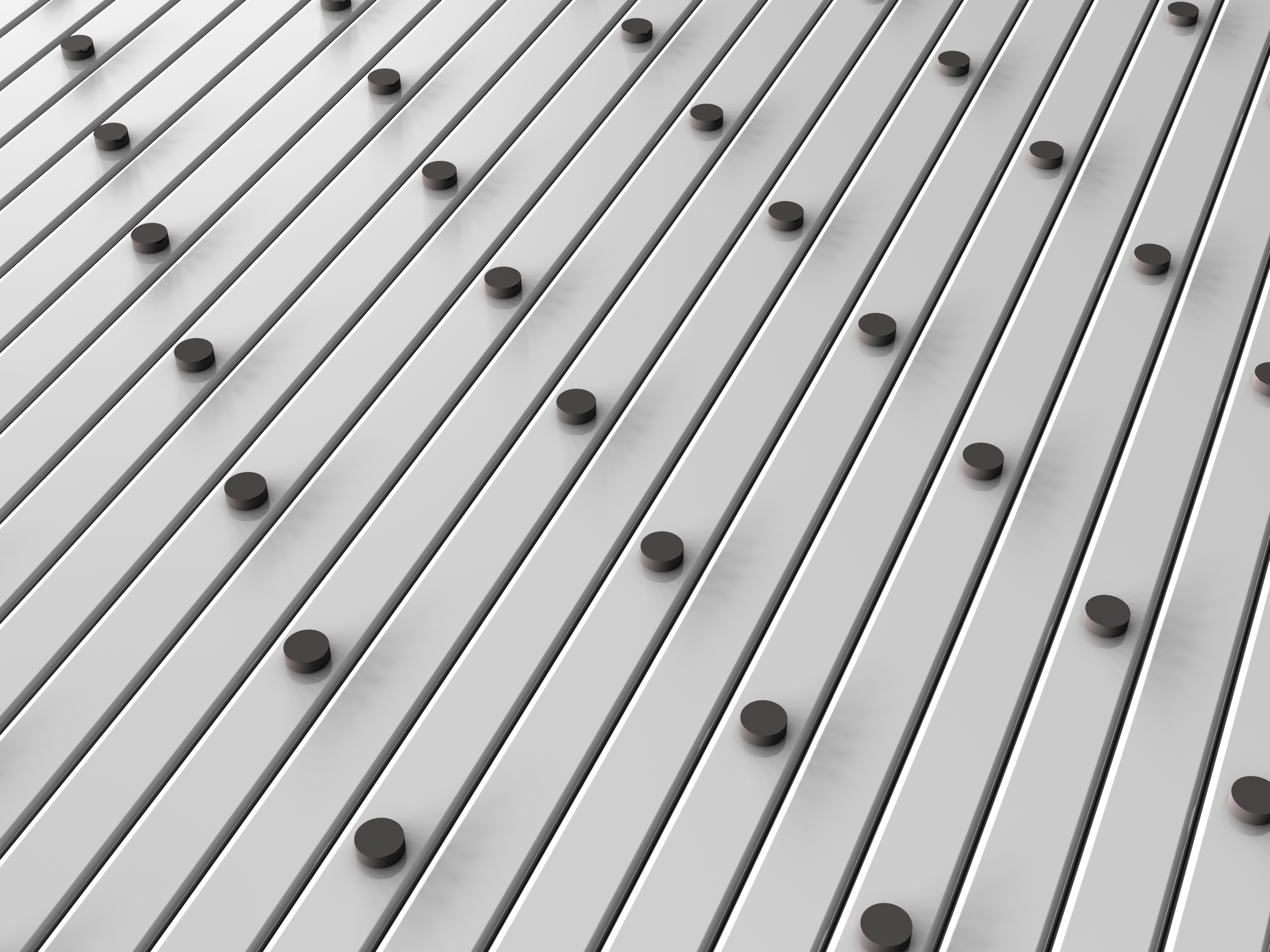}
        \caption{\it A detailed view of a closed skin constructed with a non-woven fabric structure.}
        \label{fig:00detail}
    \end{subfigure}
    \hfill  
    \begin{subfigure}[t]{0.32\textwidth}       \includegraphics[width=1.0\textwidth]{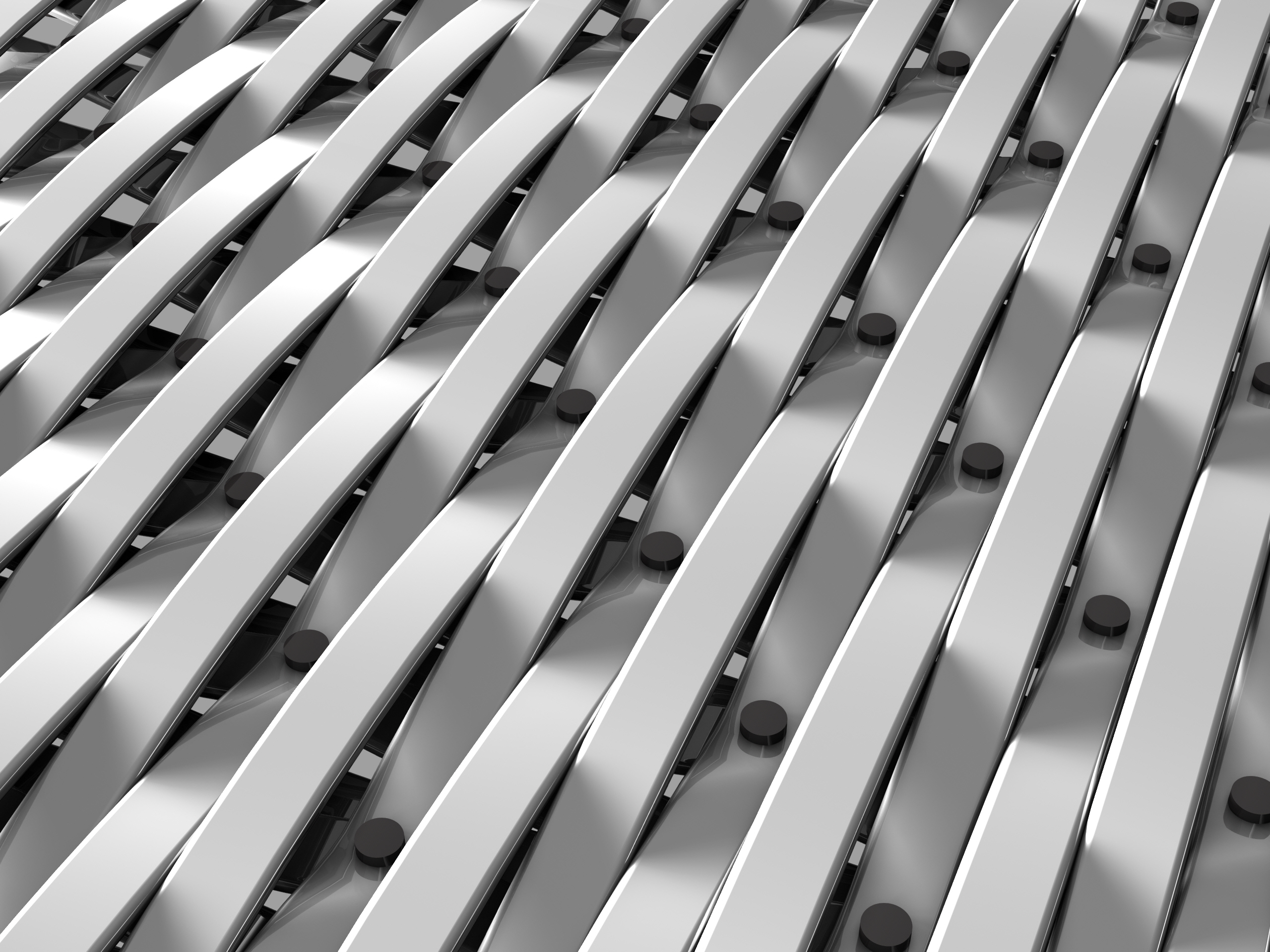}
        \caption{\it A detailed view of an open skin constructed with the same non-woven fabric structure.}
        \label{fig:01detail}
    \end{subfigure}
    \hfill  
    \begin{subfigure}[t]{0.32\textwidth}
     \includegraphics[width=1.0\textwidth]{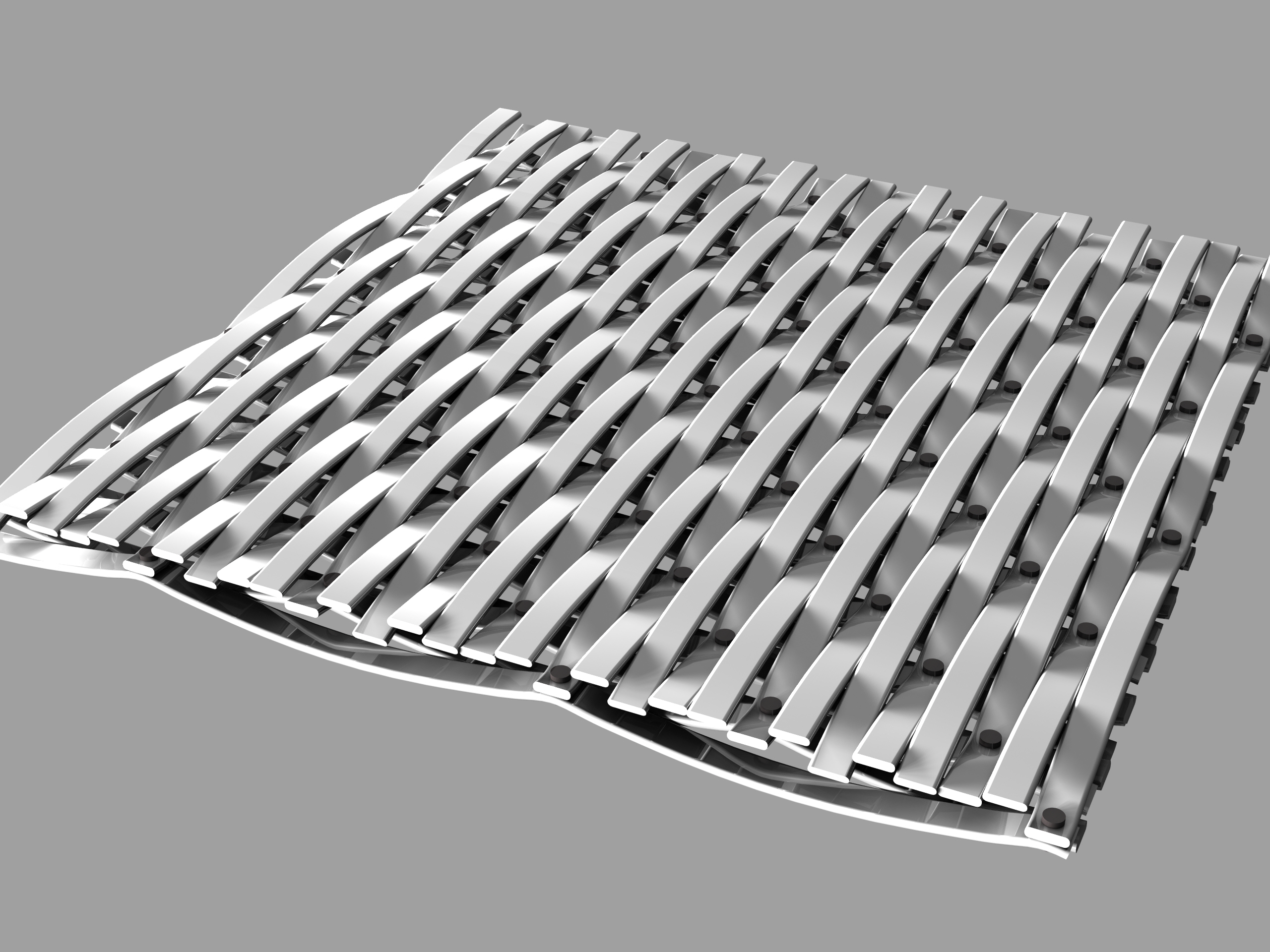}
        \caption{\it A general view of an open skin constructed with the same non-woven fabric structure.}
        \label{fig:01}
    \end{subfigure}
    \hfill 
\caption{An example of satin non-woven fabric to obtain a high amount of air exchange. This particular non-woven structure corresponds to a specific satin woven fabric, called $[7,1,3]$, which is one of the best possible among satin non-woven structures for efficiency of air exchange. The notation is explained in Section~\ref{sec:pw}}
\label{fig:main0}
\end{figure*}

\section{Introduction and Motivation} 

In the US, 50\% of all energy is consumed by buildings, and 75\% of all electricity is used for the operations of buildings \cite{perez2008}. Reduction of energy consumption in buildings can significantly reduce overall energy usage in the US. Therefore, there is a need for strategies to reduce energy consumption in buildings. Energy reduction in buildings is not an easy task since the design, construction, and operation processes of buildings are significantly different from the processes of industrial products.  For instance, cars and/or planes are designed to be mass-produced and once they are constructed, they are operated similarly. Therefore, when an energy-efficient product is developed, it is relatively simple to calculate statistically likely energy consumption of an innovative product design versus that of a former, benchmark version. 

\begin{figure*}[htpb]
    \centering 
    \begin{subfigure}[t]{0.30\textwidth}
      \includegraphics[width=1.0\textwidth]{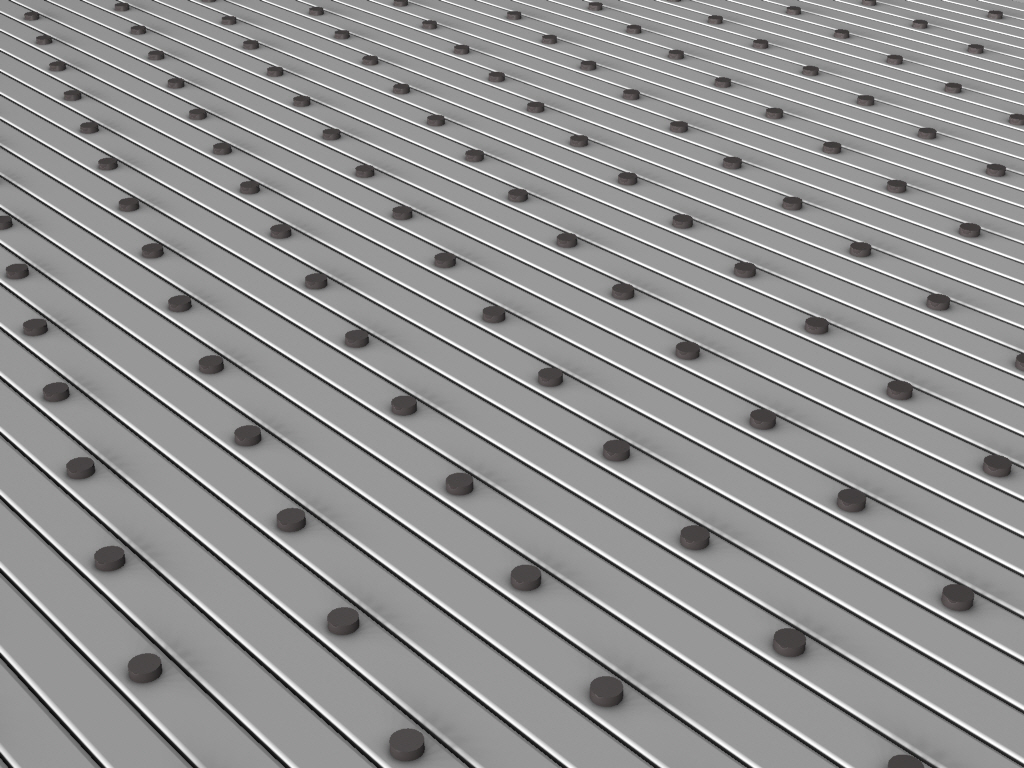}
        \caption{\it The details of the closed skin constructed with satin non-woven structure.}
        \label{fig:9-3c1}
    \end{subfigure}
    \hfill
        \begin{subfigure}[t]{0.30\textwidth}
      \includegraphics[width=1.0\textwidth]{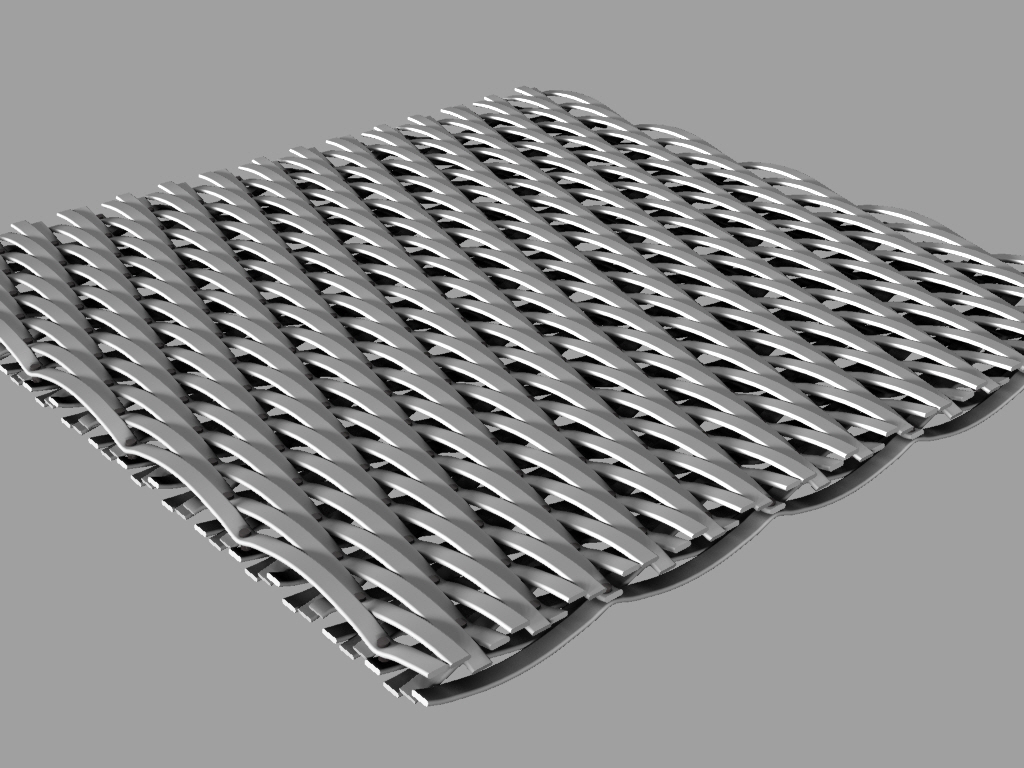}
        \caption{\it The details of the open skin constructed with satin non-woven structure.}
        \label{fig:9-3o0}
    \end{subfigure}
    \hfill   
    \begin{subfigure}[t]{0.30\textwidth}
      \includegraphics[width=1.0\textwidth]{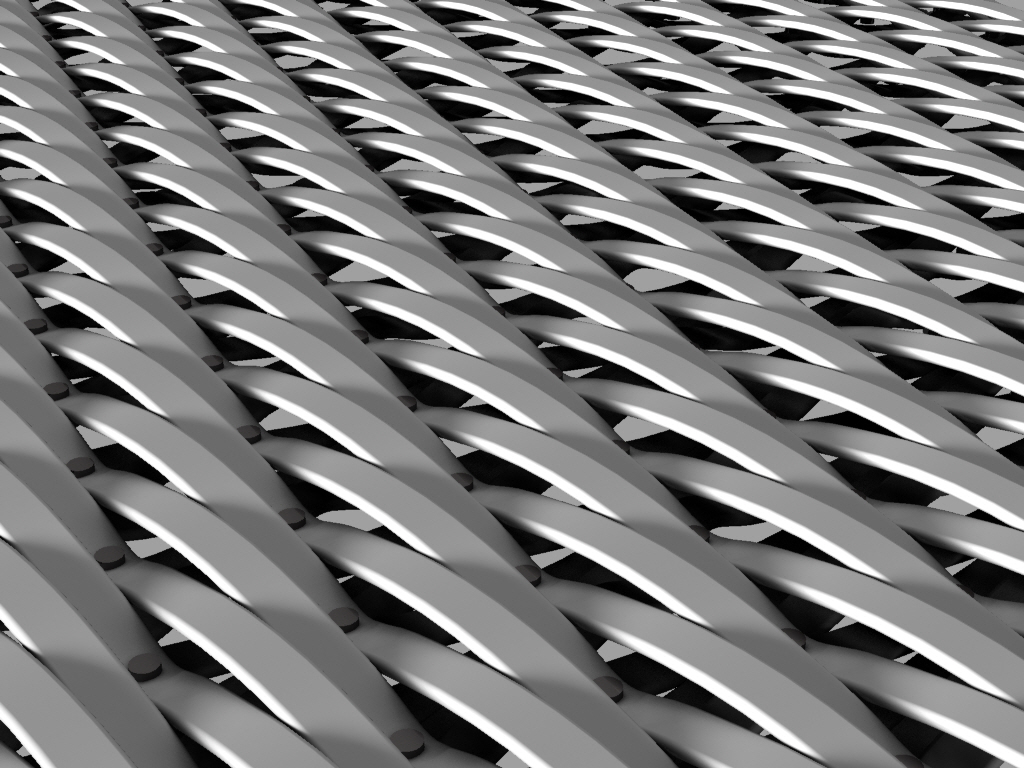}
        \caption{\it The details of the open skin constructed with satin non-woven structure.}
        \label{fig:9-3o1}
    \end{subfigure}
    \hfill
\caption{Another example of the satin non-woven structures. }
\label{fig:satin93}
\end{figure*}

\begin{figure*}[htpb]
    \centering 
    \begin{subfigure}[t]{0.30\textwidth}
      \includegraphics[width=1.0\textwidth]{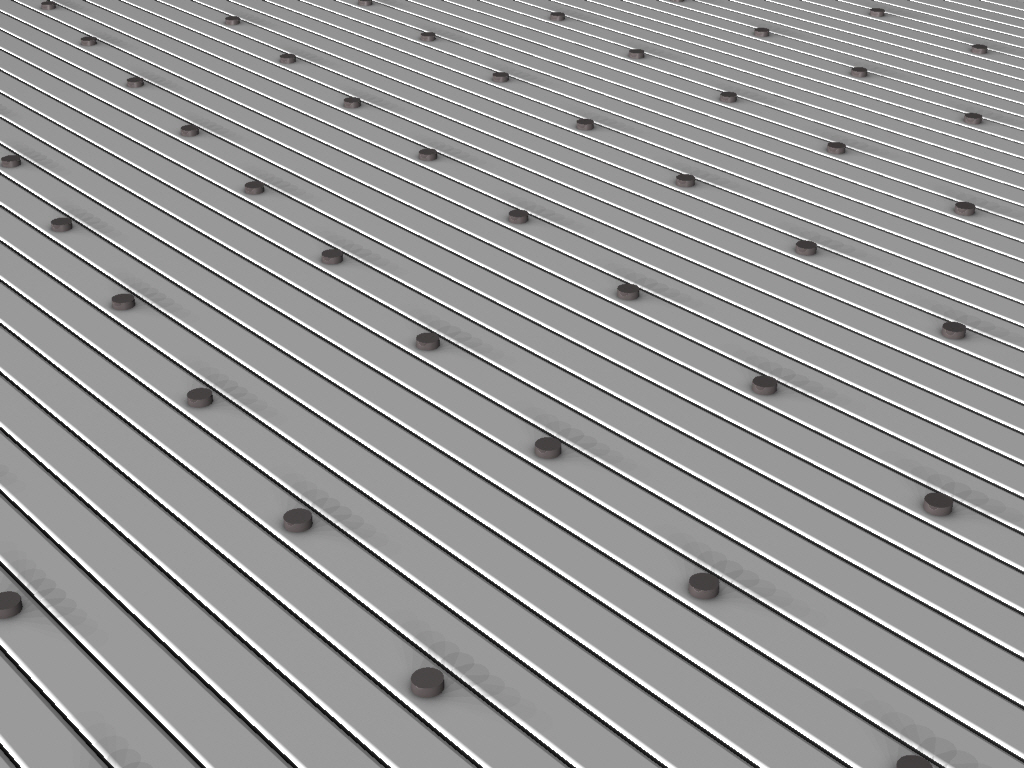}
        \caption{\it The details of the closed skin constructed with satin non-woven structure.}
        \label{fig:25-5c1}
    \end{subfigure}
    \hfill
        \begin{subfigure}[t]{0.30\textwidth}
      \includegraphics[width=1.0\textwidth]{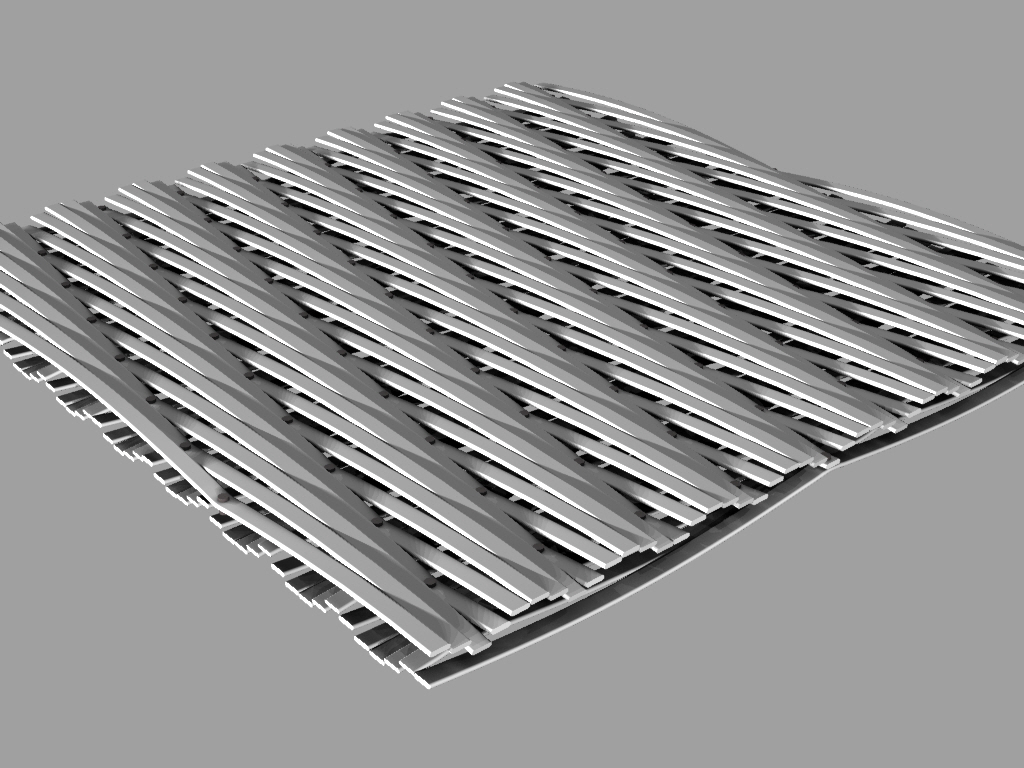}
        \caption{\it The details of the open skin constructed with satin non-woven structure.}
        \label{fig:25-5o0}
    \end{subfigure}
    \hfill   
    \begin{subfigure}[t]{0.30\textwidth}
      \includegraphics[width=1.0\textwidth]{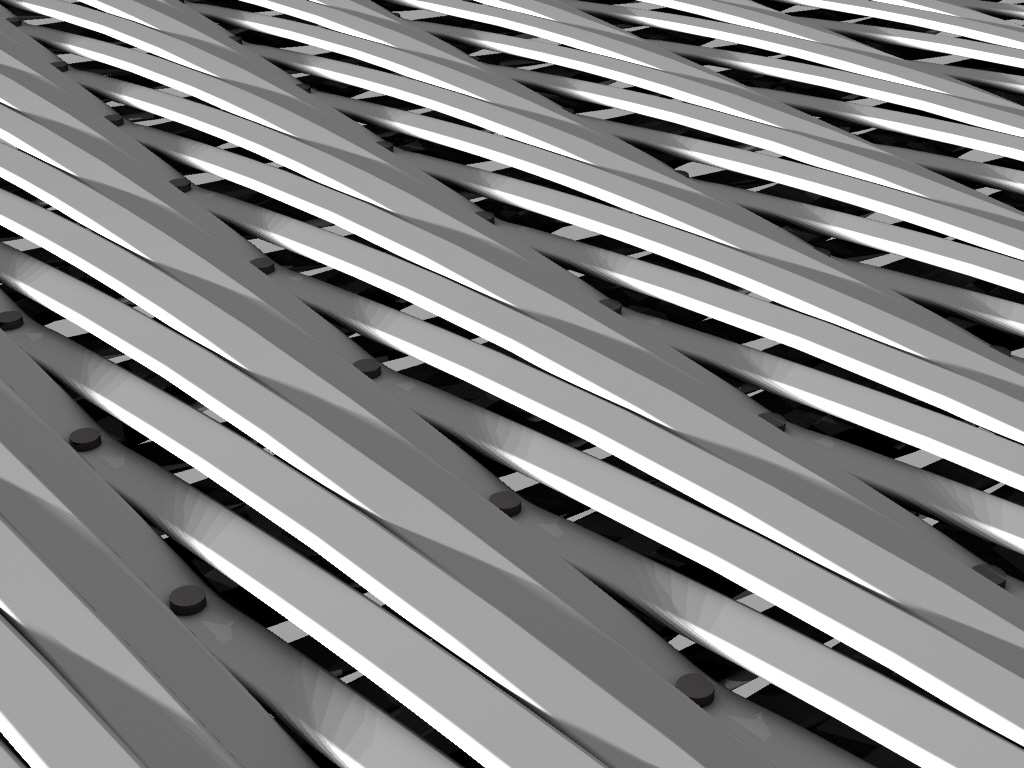}
        \caption{\it The details of the open skin constructed with satin non-woven structure.}
        \label{fig:25-5o1}
    \end{subfigure}
    \hfill
\caption{One more example of the satin non-woven structures. }
\label{fig:satin255}
\end{figure*}

Buildings, however, are typically designed, constructed, and operated as unique entities. For buildings, the best strategy is to develop modular systems that can be used for the construction of energy-efficient systems. One of the key issues to making buildings energy efficient is to make their outer skins self-regulated. We imagine the development of the type of building skins that can automatically open and close based on environmental factors similar to tiny pores called {\it “stomata’’}, which are located on the epidermis of plant leaves.  These tiny pores play a key role in transpiration by opening and closing in response to light. There is, therefore, a need to develop modules that can play the role of stomata for building skins. 

\begin{figure*}[ht]
\begin{center}
\begin{tabular}{ccccc}
\includegraphics[width=2.05in]{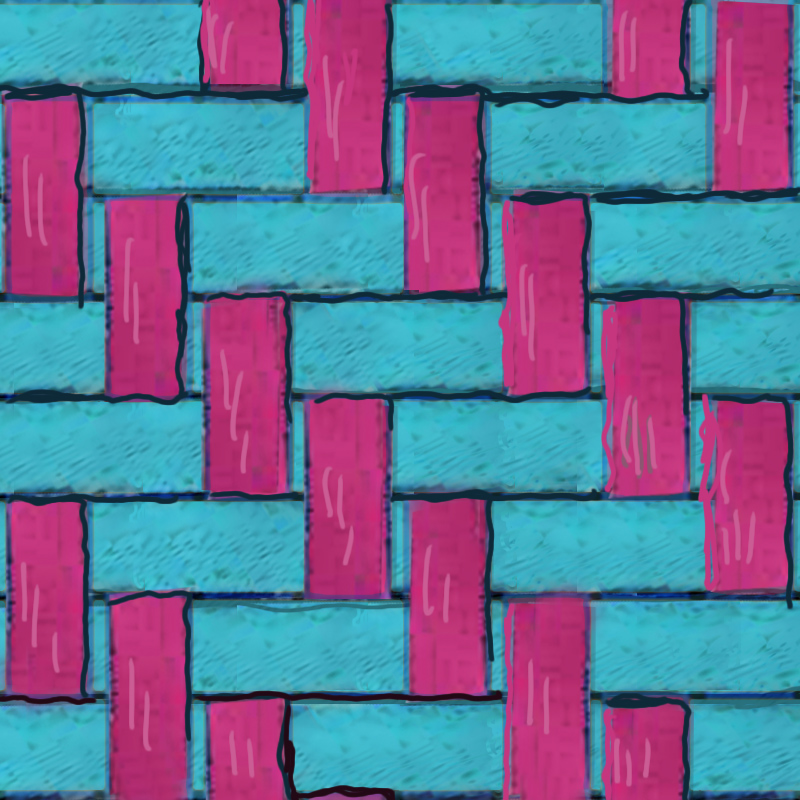}&
\includegraphics[width=2.05in]{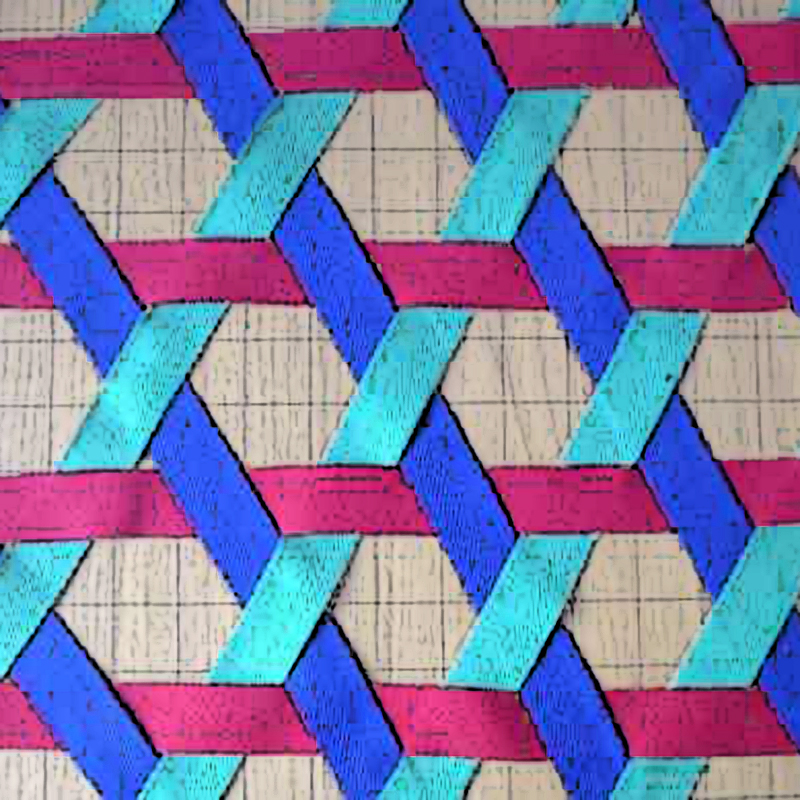}&
\includegraphics[width=2.05in]{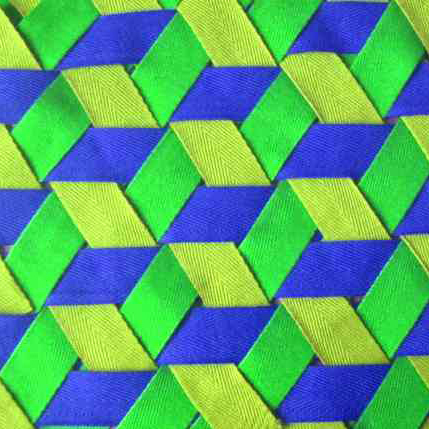}\\
\small Biaxial (2-way) Twill Fabric  & 
\small Triaxial (3-way) Sparse Fabric  & 
\small Triaxial (3-way) Dense Fabric   \\
\end{tabular}
\end{center}
 \vspace{-0.1in}
 \caption{\it Examples of 2-way 2-fold and 3-way 2-fold fabrics. 2-way, i.e. biaxial, fabrics are always dense. On the other hand, 3-way, i.e. triaxial, fabrics with straight yarns are usually sparse as shown in the second image. We, therefore, need a zigzag pattern to obtain a dense fabric as shown in the last image. All three fabrics in this example are 2-fold since there are only 2 layers in each intersection. }
  \label{fig:nway}
\end{figure*}

\begin{figure*}[htpb]
    \begin{subfigure}[t]{0.30\textwidth}
        \includegraphics[width=1.0\textwidth]{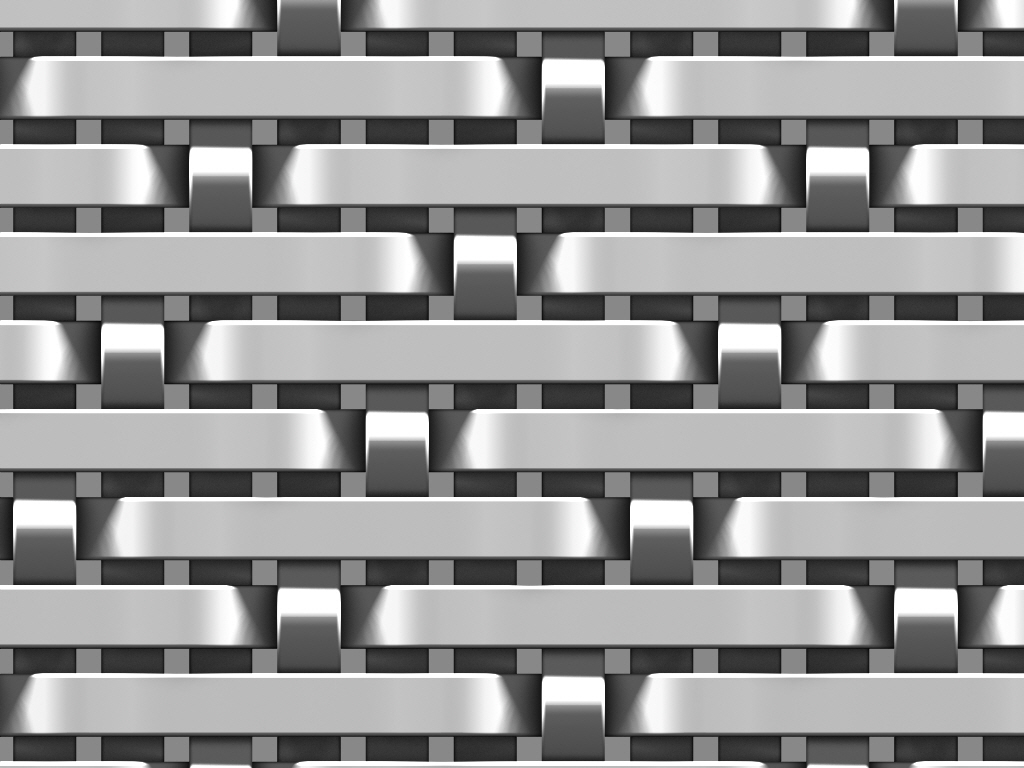}
        \caption{\it Top view of a woven structure.}
        \label{fig:a1}
    \end{subfigure}
    \hfill  
    \begin{subfigure}[t]{0.30\textwidth}
        \includegraphics[width=1.0\textwidth]{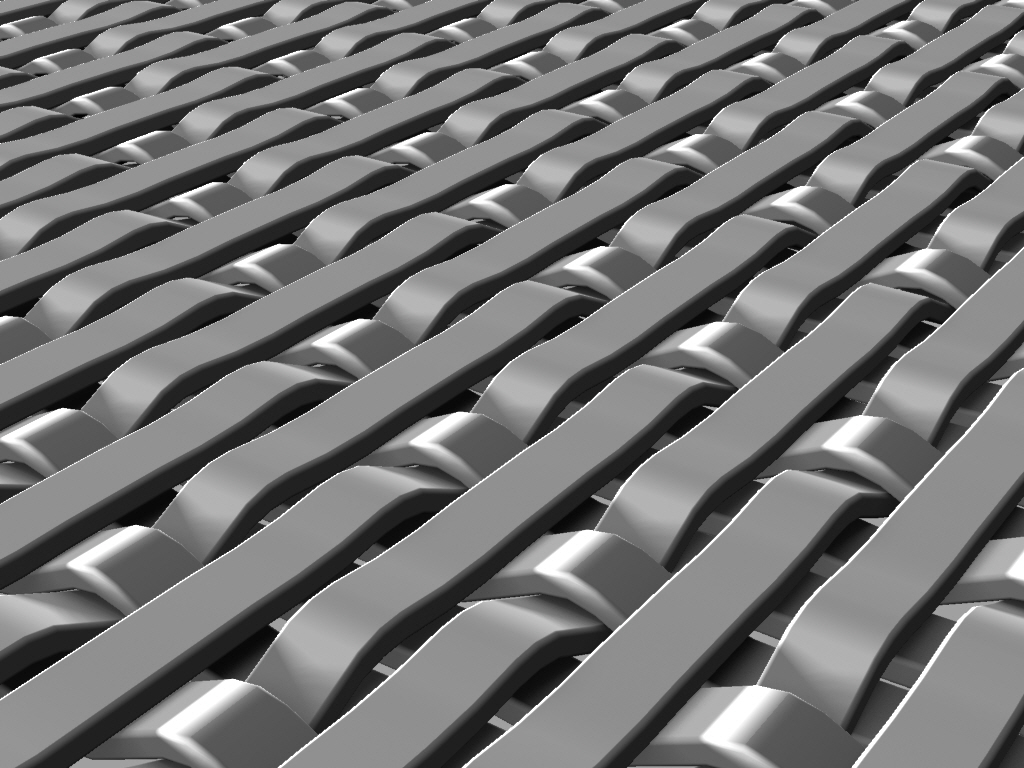}
        \caption{\it Perspective view of the center part of the same woven structure.}
        \label{fig:a3}
    \end{subfigure}
    \hfill
    \begin{subfigure}[t]{0.30\textwidth}
        \includegraphics[width=1.0\textwidth]{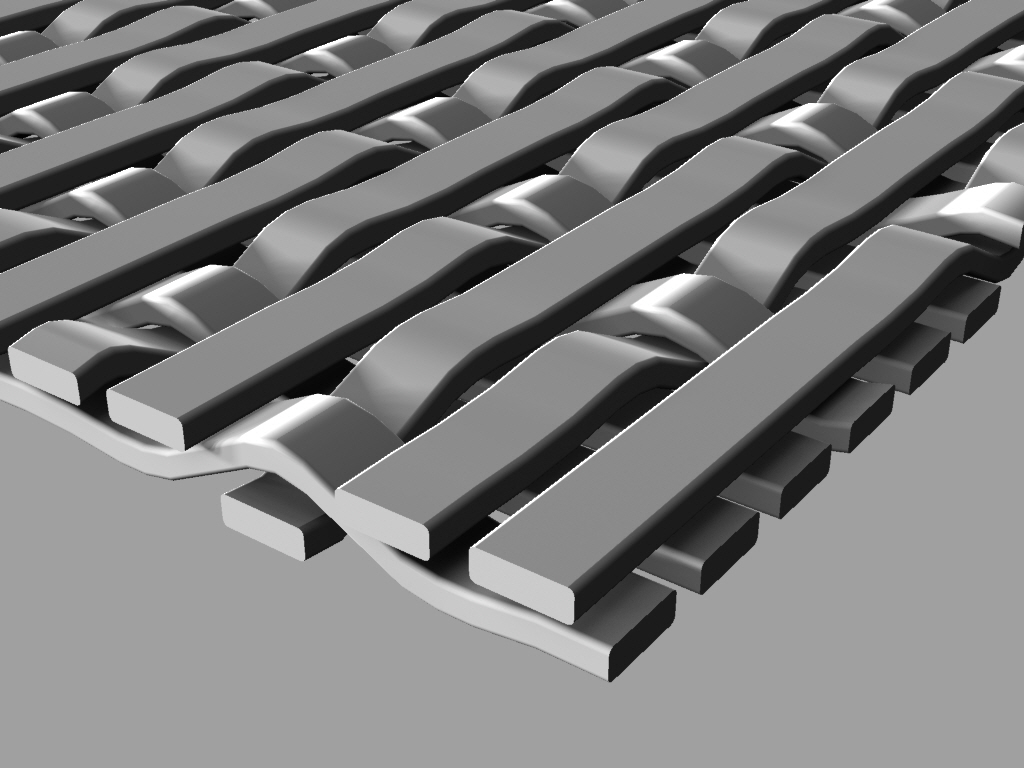}
        \caption{\it Perspective view of the side of the same woven structure.}
        \label{fig:a4}
    \end{subfigure}
    \hfill
\caption{An example that demonstrates the woven structures cannot completely shut down to avoid self-intersection if strips have non-zero thickness. The holes have to be larger for thicker strips as shown here.}
\label{fig:woven}
\end{figure*}

To play the role of stomata in buildings any self-regulatory module should (1) completely shut down when closed, and (2) allow maximum amount of air exchange when opened.  We have observed that if we replace yarns in woven or non-woven fabrics with bi-material strips we can automatically open and close modules. We have also observed that woven structures cannot completely be closed since strips must have a non-zero thickness as shown in Figure~\ref{fig:woven}. 

The advantage of non-woven structures over woven structures is that they can theoretically guarantee complete shutdown as shown in Figure~\ref{fig:main0}.  In this paper, we, further, demonstrate that satin non-woven fabrics can potentially play the role of stomata by providing maximum exchange when opened. More examples of satin non-woven structures are shown in Figures~\ref{fig:satin93}, and \ref{fig:satin255}.

Non-woven fabrics, although not well known in the research community, have been around since 1930's through a significant number of patents that focus on the engineering construction of non-woven textile structures such as \cite{brewster1934,powel1947,clayton1955,graham1956,dupre1960,allison1967}. For a comparative analysis of early works on non-woven fabrics see \cite{casper1975,purdy1983}. Non-woven fabrics have been replacing woven materials in many industrial applications from engineering to medicine since they are more reliable and they can provide better behavior for industrial applications \cite{curtis1987,balogh2015}.

Despite such a significant amount of engineering work, we could not find any work for the formalization of non-woven fabrics as mathematical/combinatorial structures. This is not surprising since the first combinatorial formalization of woven fabrics was not introduced until the 1980s \cite{Grunbaum80} although 2-way 2-fold woven fabrics such as plain, twill, and satin have been around at least five millennia \cite{horrocks2000} (See Figure~\ref{fig:nway} for examples of 2-way, 3-way and 2-fold fabrics). In this work, we offer a similar formalization that can describe all possible 2-way 2-fold genus-1 non-woven fabrics. Our work is, in fact, based on combinatorial formalization provided for woven fabrics by Grunbaum and Shephard \cite{Grunbaum80} in the early 1980's.

This combinatorial formalization for non-woven fabrics is crucial in the identification of the special subset of 2-way 2-fold genus-1 non-woven fabrics that can provide maximum air exchange when opened. In Section \ref{sec:pw}, we provide a classification of woven fabrics, and In Section \ref{sec:nwf}, we provide our classification of now-woven fabrics. Based on this classification, we identify a combinatorial structure that can define a special subset that provides maximum exchange when opened. 

\subsection{Contributions} 

In this paper, we have a series of contributions: 
\begin{enumerate}
    \item We have shown that by replacing yarns with bi-material strips we can obtain self-regulatory stomata-like behavior with both woven and non-woven fabrics. 
    \item We have shown that non-woven structures are preferable to woven structures:
    \begin{itemize}
        \item Woven structures cannot completely be closed. On the other hand, non-woven structures can completely shut down when they are closed.
        \item   Woven structures can become loose when they are opened. On the other hand, non-woven structures can still stay intact when they are opened. 
    \end{itemize}
    \item We have developed a classification of non-woven structures based on the combinatorial classification of corresponding woven structures.
    \item We have shown that the property of ``hanging togetherness'' in woven fabrics also applies to non-woven structures and it guarantees that non-woven structures are completely connected. 
    \item We have identified a specific subset of satin non-woven structures that can provide maximum air exchange. This subset is also guaranteed to be connected by having hanging together property.

\end{enumerate}

\section{Previous Work}
\label{sec:pw}

This paper is inspired by woven fabrics that can effectively provide a reasonable amount of air exchange through tiny openings caused by the thickness of yarns or strips as shown in Figure~\ref{fig:woven}. 
Most common woven fabrics are {\it biaxial}, or 2-way and 2-fold, which consist of two strands, i.e. 2-way that are called warp and weft. The word 2-fold originates from the behavior of the weft strands that pass over and under warp strands, which corresponds to the thickness of the fabric. The mathematics behind such 2-way, 2-fold woven fabrics such as plain, satin, and twill, are first formally investigated by Grunbaum and Shephard \cite{Grunbaum80}. They also coined the word, isonemal fabrics to describe fabrics that have a transitive symmetry group on the strands of the fabric \cite{Grunbaum88}. One of the important concepts introduced by Grunbaum and Shephard is {\it hang-together} or {\it fall-apart}. They pointed out that weaving patterns that look perfectly reasonable may not produce interlacements that {\it hang-together}. In such patterns, some interlaced warp and weft threads may not be interlaced with the rest of the fabric if the fabric is woven. Such a fabric would come apart in pieces. Determining whether or not a pattern is {\it hang-together} or {\it fall-apart} cannot be identified by simple visual inspection of the pattern. Several procedures have been developed to identify the {\it hang-together} property of a weaving pattern. If a weaving pattern is hanging together, it is then called a fabric. Grunbaum and Shephard provided images of all non-twill isonemal fabrics (i.e. hanging-together weaves) of periods up to 17 \cite{Grunbaum85,Grunbaum86}. Their papers were followed by several others giving algorithms for determining whether or not a fabric hangs together \cite{Griswold2004a,Clapham80,Clapham85,Enns84,Delaney84}.
Roth \cite{Roth1993}, Thomas \cite{Thomas2010,Thomas2009,Thomas2009a} and Zelinka \cite{Zelinka1983,Zelinka1984} and Griswold \cite{Griswold2004b,Griswold2004c} theoretically and practically investigated symmetry and other properties of isonemal fabrics.

It was recently demonstrated that any surface can be covered with plain woven structures
\cite{akleman2009cyclic}, twill woven structures \cite{akleman2011cyclic}, and it is also possible to cover any surface with a single cycle \cite{xing2010single}. The recent mathematical work further demonstrated that surfaces can be covered by any type of woven structure \cite{akleman2020topologically,akleman2015extended}. 

\begin{figure*}[ht]
\begin{center}
\begin{tabular}{ccccc}
\includegraphics[width=1.15in]{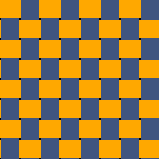}&
\includegraphics[width=1.15in]{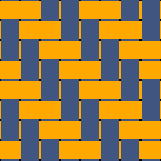}&
\includegraphics[width=1.15in]{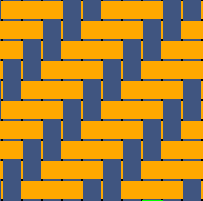}&
\includegraphics[width=1.15in]{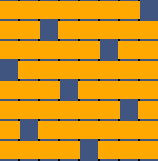}&
\includegraphics[width=1.15in]{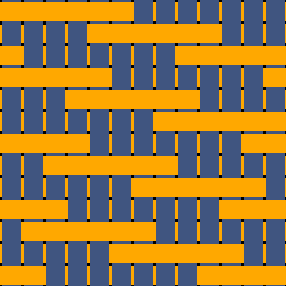}\\
\small Plain: $[1,1,1]$  & 
\small Twill: $[2,2,1]$  & 
\small Twill: $[3,2,4]$ &
\small Satin: $[7,1,3]$  &
\small Unnamed: $[6,7,4]$ \\
\end{tabular}
\end{center}
\vspace{-0.1in}
 \caption{\it Examples of textile weaving patterns.}
  \label{fig:example1}
\end{figure*}

In this paper, we are interested in the non-woven generalization of certain types of isonemal fabrics that are called genus-1 by Grunbaum and Shephard \cite{Grunbaum86}. Genus-1 means that each row with length $n$ is obtained from the row above it by a shift of $s$ units to the right, for some fixed value of the parameter $s$. They call these fabrics  $(n, s)$-fabrics where $n$ is the period of the fabric and $n$ and $s$ are relatively prime. The genus-1 family of fabrics includes the most common fabrics: plain, satins, and twills. If $s=1$ the fabric is a twill and $s^2 \equiv 1 \mbox{mod} n $ corresponds to satin for the fabrics that contain exactly one black square in every row. In Grunbaum and Shephard's notation plain weaving in Figure~\ref{fig:example1}(a) is called $(2, 1)$-fabric, twill weaves in Figure~\ref{fig:example1}(b) and~(c) are called $(4, 1)$ and $(5,4)$-fabrics respectively, satin weaving in Figure~\ref{fig:example1}(d) is called $(8, 3)$-fabric and finally the unnamed weave in Figure~\ref{fig:example1}(e) now has  a name and is called $(13,4)$-fabric. Genus-1 fabrics not only include satins and twills but also are the only isonemal fabrics that exist for every integer $n$. The other isonemal fabrics do not exist for odd $n$ integer values \cite{Grunbaum86}.

One problem with Grunbaum and Shephard's $(n, s)$ notation is that it ignores the initial row pattern. Chen et al. showed that it is useful to express the initial pattern by two integers $a$ and $b$, where $a$ is the number of up-crossings, and $b$ is the number of down-crossings where $a+b=n$ and an additional integer $s$ still denotes the shift introduced in adjacent rows \cite{chen2010}. Any such weaving pattern can be expressed by a triple $[a,b,s]$.  The Figure~\ref{fig:example1} also shows a basic block of a biaxial weaving structure and the role of these three integers, $a >0 $, $b > 0$ and $0 < s < a+b$ \footnote{The value of $s$ can be any integer, however, $[a,b,s]$ and $[a,b,s+ k(a+b)]$ are equivalent since $s+ k(a+b)) \equiv s \mbox{mod} (a+b)$. Therefore, we assume that the value of the $s$ is between $0$ and $a+b$.}. 
Figure~\ref{fig:example1} shows four examples of biaxial weaving structures that can be described by the $[a,b,s]$ notation. Using these notations it is possible to name each weaving structure uniquely. 

The notation $[a,b,s]$ can represent a fabric that can fall apart. However, unlike a general isonemal weaving case, it is easy to avoid the fabric falling apart. A genus-1 weaving pattern produces a fabric structure that hangs together if and only if  $a+b$ and $s$ are relatively prime (i.e. $gcd(a+b,s)=1$). In other words, if the greatest common divider of $a+b$ and $s$ is not $1$, the resulting structure is not a fabric since it cannot hang together and, therefore, must be avoided. This provides sufficient background for woven fabrics. We now are ready to discuss non-woven structures. 

\section{Non-Woven Fabrics}
\label{sec:nwf}

Non-woven fabrics are obtained by bonding fibers to each other using some sort of adhesive. We note that a non-woven structure can also be described by just defining where adhesives are applied. For instance, the two colored grids in Figure~\ref{fig:example1} can be considered representations of non-woven structures, if we assume that the adhesives are applied to the blue squares. Under this assumption, we can represent non-woven structures also with three numbers as $[a,b,s]$ where $a$ is the length of the free region, $b$ is the length of the bonded region given that $a+b=n$, and the additional integer $s$ still denotes the shift introduced in adjacent rows \cite{chen2010}.  

\begin{figure*}[htpb]
    \begin{subfigure}[t]{0.30\textwidth}
        \includegraphics[width=1.0\textwidth]{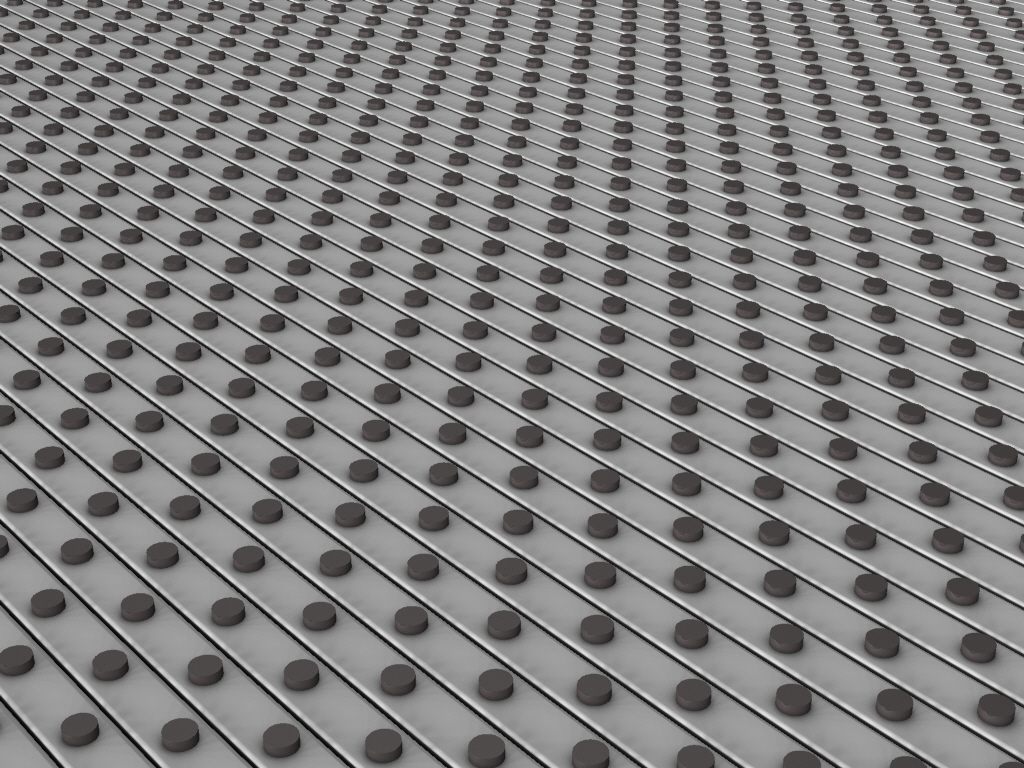}
        \caption{\it A detailed view of the closed skin constructed with plain non-woven structure.}
        \label{fig:11c1}
    \end{subfigure}
    \hfill  
    \begin{subfigure}[t]{0.30\textwidth}       \includegraphics[width=1.0\textwidth]{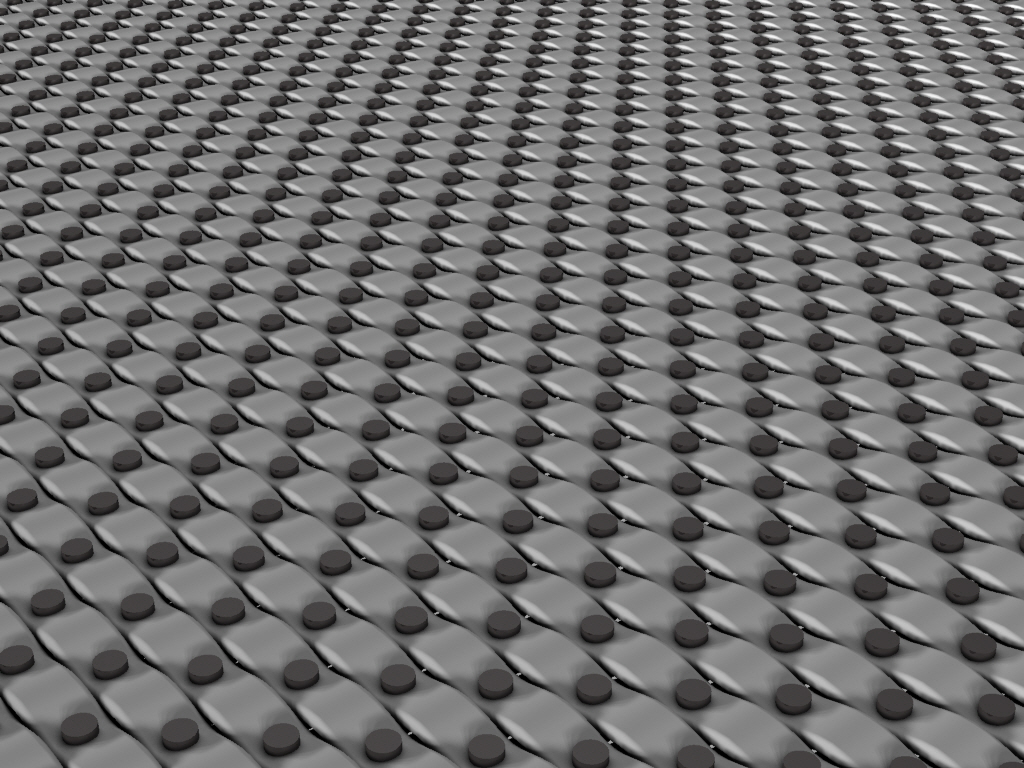}
        \caption{\it A detailed view of the open skin constructed with plain non-woven structure.}
        \label{fig:11o1}
    \end{subfigure}
    \hfill 
    \begin{subfigure}[t]{0.30\textwidth}
     \includegraphics[width=1.0\textwidth]{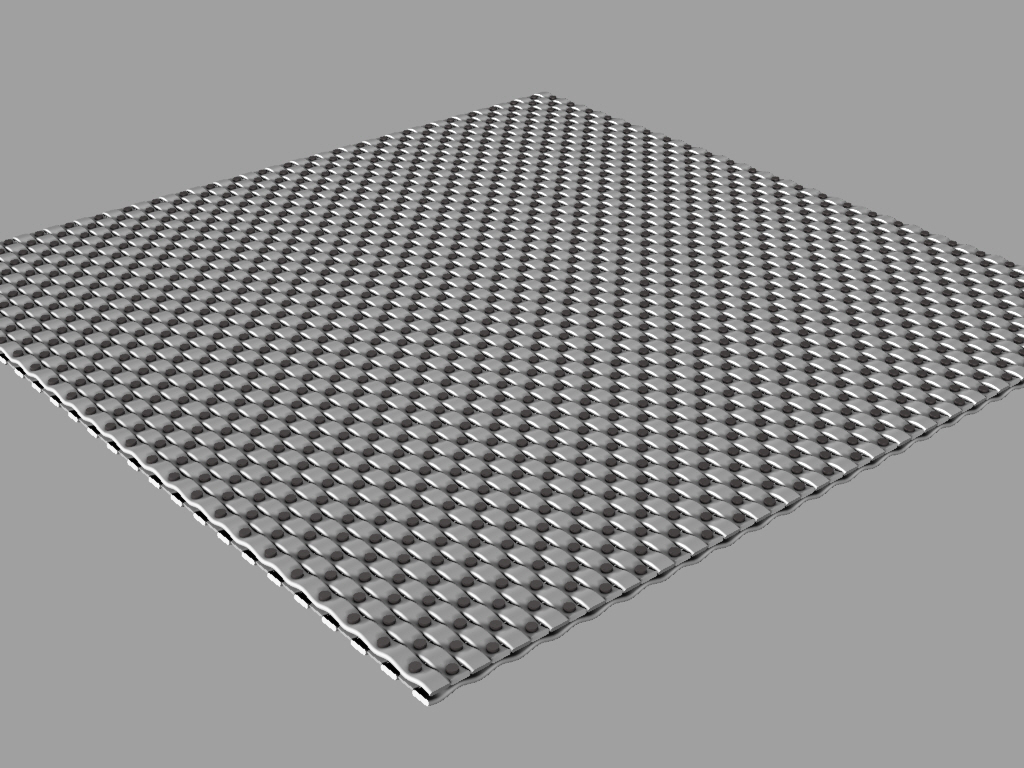}
        \caption{\it An open skin constructed with a plain non-woven structure.}
        \label{fig:11o0}
    \end{subfigure}
    \hfill 
\caption{An example that demonstrates the plain non-woven structures cannot provide sufficient openings for air exchange. }
\label{fig:plain}
\end{figure*}

One of the key issues with non-woven fabrics is to reduce the bonded region as much as possible. Since we want to keep the resulting structure connected, $b$ should be larger than $0$. Since the smallest integer larger than $0$ is $1$, it is better to keep $b$ always equal to $1$, which makes $a=n-1$. Therefore, we are interested in non-woven fabrics only in the form of $[n-1,1,s]$. Figure~\ref{fig:plain} shows a non-woven version of the simplest woven fabrics, $[1,1,1]$, called plain weaving. Such plain non-woven fabrics are frequently used in the construction of many types of materials and, therefore, this structure probably looks familiar to many people. However, as it can easily be seen plain is not a good structure to provide air exchange since there is not enough room for strands to curve out.

\begin{figure*}[htpb]
    \begin{subfigure}[t]{0.22\textwidth}
       \includegraphics[width=1.0\textwidth]{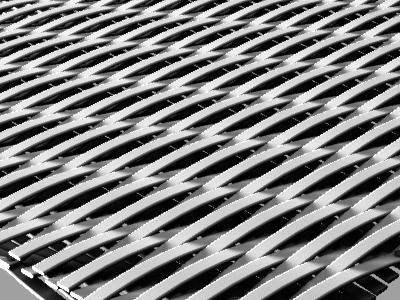}
        \caption{\it The details of front-side view of $[14,1,7]$ non-woven structure in open position.}
        \label{fig:14-1-7}
    \end{subfigure}
    \hfill  
    \begin{subfigure}[t]{0.22\textwidth}
      \includegraphics[width=1.0\textwidth]{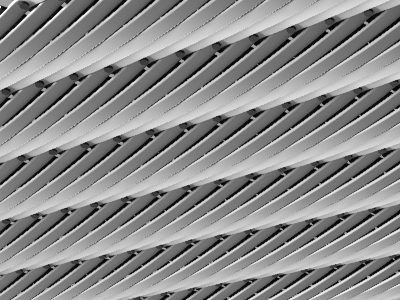}
        \caption{\it The details of back-side view of $[14,1,7]$ non-woven structure in open position.}
        \label{fig:14-1-7b}
    \end{subfigure}
    \hfill  
     \begin{subfigure}[t]{0.22\textwidth}
       \includegraphics[width=1.0\textwidth]{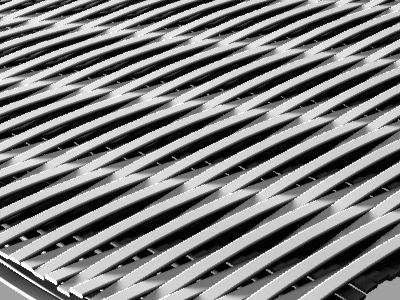}
        \caption{\it The details of front-side view of $[24,1,12]$ non-woven structure in open position.}
        \label{fig:24-1-12}
    \end{subfigure}
    \hfill  
    \begin{subfigure}[t]{0.22\textwidth}
      \includegraphics[width=1.0\textwidth]{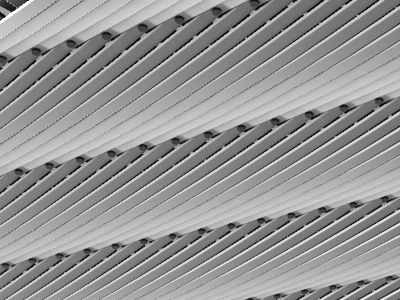}
        \caption{\it The details of back-side view of $[24,1,12]$ non-woven structure in open position.}
        \label{fig:24-1-12b}
    \end{subfigure}
    \hfill
\caption{An example that demonstrates the $[2n,1,n]$ non-woven structures cannot provide sufficient openings for air exchange in the backside. }
\label{fig:problems}
\end{figure*}

In order to get maximum air exchange, for a given $n$, $s$ has to be chosen in such a way that there should be a maximum shift between two consecutive warps (or wefts). In other words, $s$ must be the closest integer to $(n-1)/2$, i.e. $s = \lfloor (n-1)/2 + 0.5 \rfloor$. If we choose an even number, the notation simplifies to $[2s,1,s]$. This particular structure looks promising as a first impression, however, the other (bottom/weft) side crates problems since the other side (bottom/weft) is not guaranteed to be the same as the warp side (See Figure~\ref{fig:problems} for the problems in back-side). 

It can be proven that when the top strands follow $[n-1,1,s_1]$ non-woven pattern, the bottom strands will follow $[n-1,1,s_2]$ non-woven pattern. It can also be shown that $s_1$ and $s_2$ are not independent and their dependency can be described as $s_1 \; s_2 \equiv 1 \mbox{mod} n $. The best solution to maximize both $s_1$ and $s_2$  is, therefore, to choose the same integer value for  $s_1$ and $s_2$. Another acceptable solution is to choose numbers for them that only differ by one $|s_1 - s_2 | =1$. Assuming these constraints, we obtain either 
$s^2 = n+1$ or  $s^2 - s = n+1$. These two equations give us an approximation for $s$ as $s \approx \sqrt {n+1}$. It is interesting to note that these settings correspond to satin woven fabric structures. Therefore, it is better to call the optimized solutions satin non-woven fabrics.

\begin{figure*}[htpb]
    \centering 
    \begin{subfigure}[t]{0.30\textwidth}
      \includegraphics[width=1.0\textwidth]{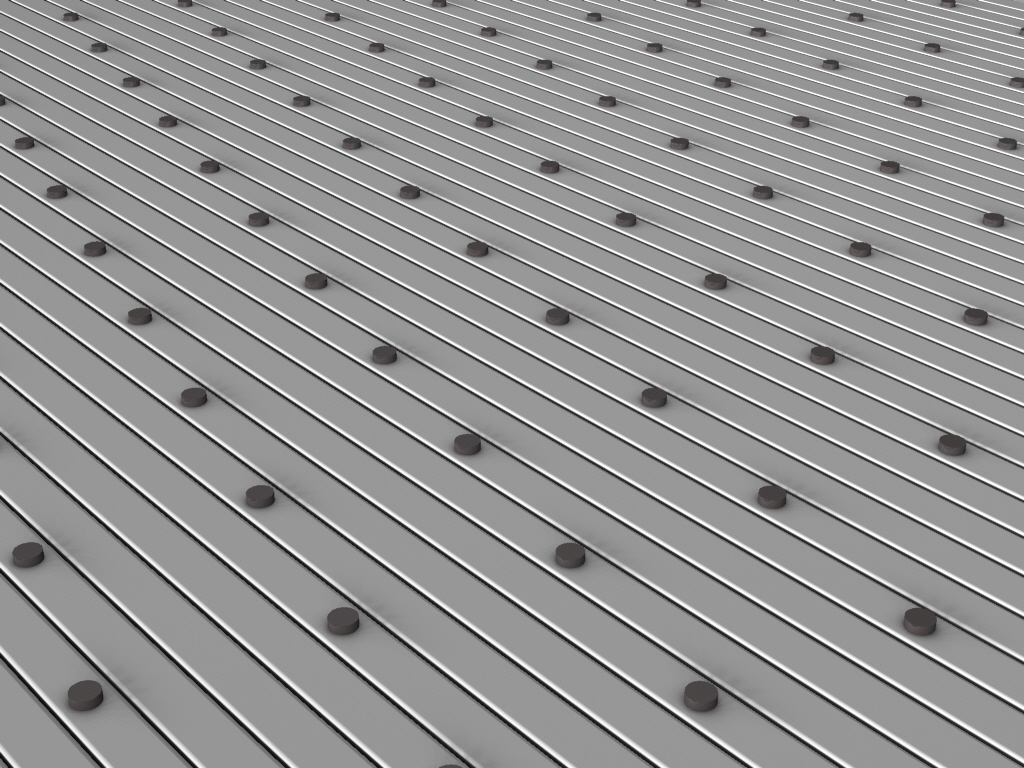}
        \caption{\it The details of the closed skin constructed with satin non-woven structure.}
        \label{fig:164c1}
    \end{subfigure}
    \hfill   
    \begin{subfigure}[t]{0.30\textwidth}
       \includegraphics[width=1.0\textwidth]{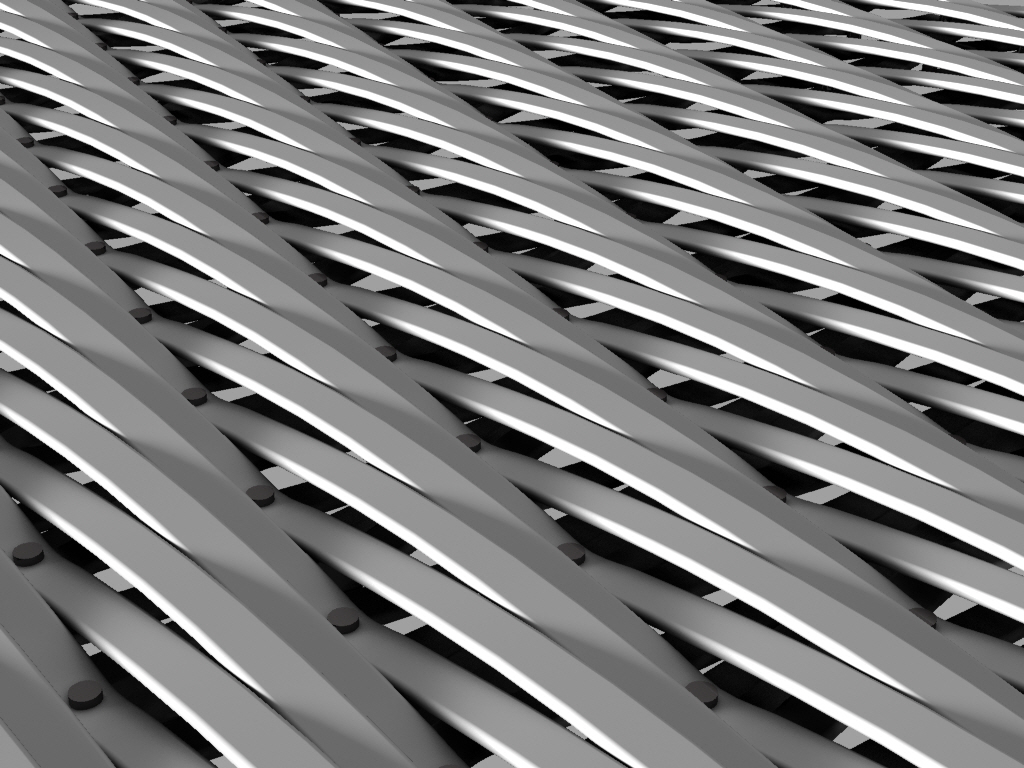}
        \caption{\it The details of the open skin constructed with satin non-woven structure.}
        \label{fig:164o1}
    \end{subfigure}
    \hfill 
    \begin{subfigure}[t]{0.30\textwidth}       \includegraphics[width=1.0\textwidth]{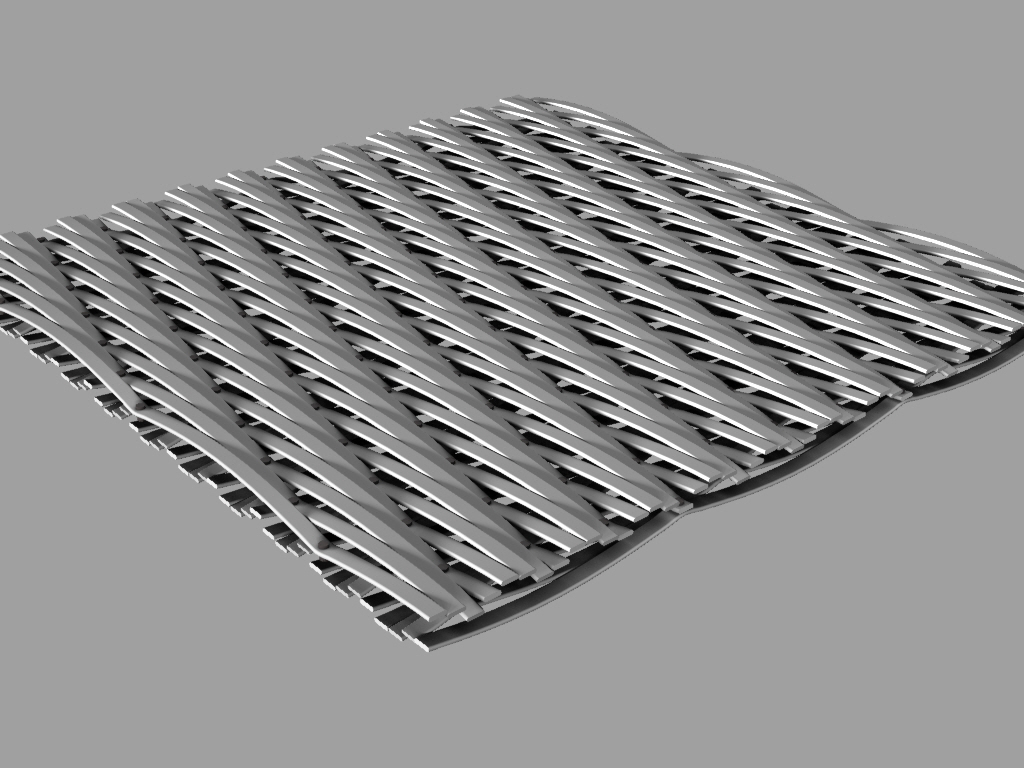}
        \caption{\it The open skin constructed with satin non-woven structure.}
        \label{fig:164o0}
    \end{subfigure}
    \hfill
\caption{An example that demonstrates the satin non-woven structures can provide more sufficient openings than plain structures. However, the air exchange can never get as good as $(2n+1,1,n)$ non-woven structures. }
\label{fig:satin164}
\end{figure*}

One of the important issues is the identification of the material properties of strips such that they can automatically open and close based on some environmental changes such as temperature or humidity. We first considered single-material strips and found out that they do not provide automatic opening and closing unless we use an additional scaffold base. We, therefore, analyzed bi-material strips based on the inspiration coming from bi-metallic strips that have been used in many applications such as thermostatic switches to obtain self-regulating behavior \cite{dubilier1942,brennan1951}.

\section{Bi-Material Strips}

\begin{figure*}[htpb]
    \centering 
    \begin{subfigure}[t]{0.45\textwidth}
     \includegraphics[width=1.0\textwidth]{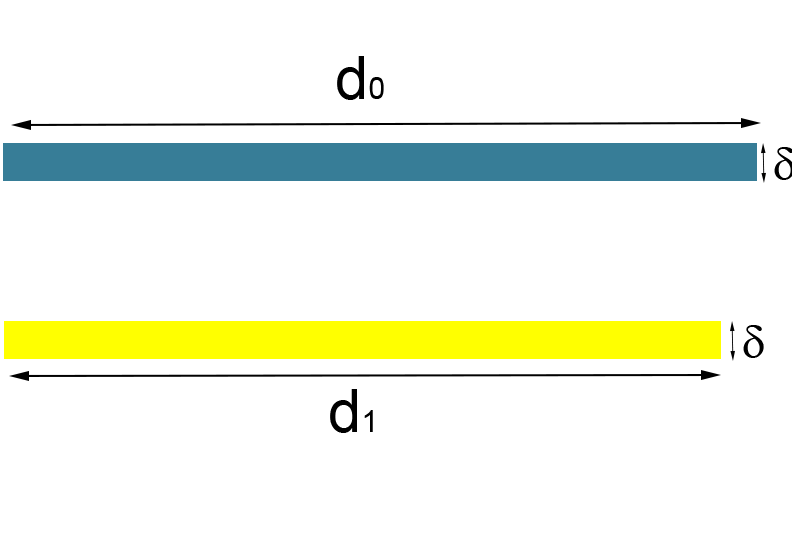}
        \caption{\it Two strips with different lengths that are not bonded together. }
        \label{fig:c0}
    \end{subfigure}
    \hfill    
    \begin{subfigure}[t]{0.45\textwidth}
      \includegraphics[width=1.0\textwidth]{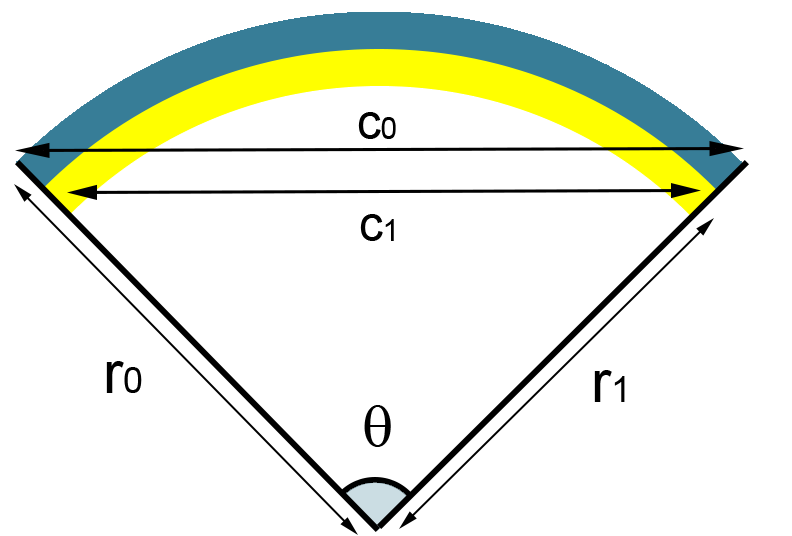}
        \caption{\it When the two strips with different lengths are bonded together they become curved.}
        \label{fig:c1}
    \end{subfigure}
    \hfill
\caption{An example that demonstrates how two strips with different lengths become curved when they are bonded together. }
\label{fig:curved}
\end{figure*}

In this section, we demonstrate that bi-material strips, which are shown in Figure~\ref{fig:c1} as blue and yellow rectangles that are bonded together, can be useful to automatically bend warp and wefts based on different amounts of expansions of the yellow and blue stripes. The main idea is that the difference in thermal expansions of two strips causes the bonded bi-material strip to bend as shown in Figure~\ref{fig:curved}.  In this paper, we ignore the cause or the nature of the expansion. Without loss of generalization and for the simplicity of discussion, we assume that the expansions are caused only by changes in temperature and can be explained by multiplication with an expansion ratio. 

Now, let both stripes have the thickness denoted by $\delta$. Let $d=d'_0=d'_1$ denote the original length of blue and yellow strips when it is not curved, and let $d_0$ and $d_1$ denote the length of curved blue and yellow strips. Without loss of generality, we can assume that the shape of the curved bi-material strip is a circle segment as shown in Figure~\ref{fig:c1} given a radius $r$ and angle $\theta$. Based on this assumption, we can write the following two equations: 
$$d_1 = r_1 = r \theta, 
\; \; \; \; \; \; \; \;\mbox{and} \; \;  \; \; \; \; \; \;
d_0 = r_0 \theta = (r + \delta ) \theta$$
Let $D$ denote the relative expansion between the two materials, which is given as the ratio between the two stripes as
$D=d_1/d_0$. Based on this, we can compute $D$ independent of $\theta$ as and we compute the radius $r=r_1$ as follows:
$$D = \frac{r \theta}{(r + \delta ) \theta} =\frac{r}{r+\delta}  
\; \; \; \; \; \; \; \;\Longrightarrow \; \; \; \; \; \; \; \;
r = \frac{\delta D}{1-D}$$
We can write this equation in terms of curvature by using the fact that the curvature $\kappa$ is inverse proportional to radius $r$. As a result, we obtain, 
$$\kappa = \frac{1-D}{\delta D}; 
\; \; \; \; \; \; \; \;\mbox{note that} \; \;  \; \; \; \; \; \;
\lim_{\delta D \rightarrow 0} \kappa \longrightarrow \infty $$
Using this key equation, we can design bi-material strips that can provide the desired curvature based on changes in the temperature.

\paragraph{Remark 1:} The denominator term $\delta D$ is useful for our applications since by making either or both of the two parameters $\delta$ and $D$, we can precisely control curvature and we can even obtain arbitrarily high curvature values. Because of the physical constraints neither one of them in practice can go to zero but we can change them simultaneously to obtain the desired curvature. 

\paragraph{Remark 2:} We want to point out that $D$ is not the thermal expansion of a single material; it is rather the relative thermal expansion, which is given as a ratio of two thermal expansions: $d_0/d$ and $d_1/d$. Using structures from phase-changing materials such as paraffin, we can theoretically obtain up to $25\%$ thermal expansions, which corresponds to $d_0/d = 1.25$. If the other material, i.e. yellow, do not extend, we get $D$ values up to $d_1/d_0 = 1/1.25 = 0.8$, of course, such a high relative thermal expansion is hard to get in practice. Fortunately, this is not a problem for our applications.  

\paragraph{Remark 3:} An important implication of this result is that we do not even need high $D$ values to obtain high curvature. For example, assume that we want to use two materials with a relative expansion ratio of $D=0.999$, and we want to obtain a radius $r=1.00 \; m$, since $\delta  = (1-D) r / D = 0.001 /0.999 m \approx 0.001 \; m = 1 \; mm$, which is not really impossible to obtain.

For designing a large opening, curvature is not the only factor. The amount of opening is the function of areas, not the curvature. Now, let $A_T$ denote the area of the isosceles triangle formed under $c_1$ shown in Figure~\ref{fig:c1}, whose length is given as  $c_1 = 2 r sin \theta $, then
$A_T = r^2 sin \theta$
and let $A_C$ denote the area of the circle bounded yellow strip and the two lines, then 
$A_C = r^2  \theta $. Let $\Delta_A$ denote the open area, which  is the difference of the areas $A_C$ and $A_T$ as 
$$\Delta_A = A_C - A_T = r^2 \; \;  (\theta - \sin \theta )$$
This is another interesting formula that demonstrates we do not need high curvature to obtain large openings since the size of the opening is directly proportional to  $r^2$ and $(\theta - \sin \theta)$. In other words, to obtain larger openings, we need to increase either (1) $r^2$  and (2) $\theta$ or both of them. 

In the previous example, let us assume that in addition to $r=1.00 \;    m$ we obtain $\theta = 0.5\;  \mbox{radian} \approx 30  \;  \mbox{degree}   $, then the opening is computed as  $\Delta_A = 0.500 - 0.479 \approx 0.02 \;  m^2$.
To visualize the area of this opening, try to mentally construct a square with sides $15 \;  cm$, which is significantly large, much more than we need.  
If $\theta = 0.14 \; \mbox{radian}$, which is approximately $10$ degree, then the opening is computed as $\Delta_A = 0.100 - 0.099 \approx 0.001 \;  m^2$
To visualize the area of this opening, try to mentally construct a square with sides $3  \; cm$, which is still significantly large. This demonstrates that it is possible to obtain the desired amount of openings in most cases based on environmental requirements for self-regulation.

\subsection{The Effect of the Back Side}

We also need to consider the effect of the back side in creating curved pieces. The back side consists of $n-1$ strips. Each strip must have a width $d/a$ when they are not curved where $a=n-1$. When they are curved the width of the blue strip becomes $d_0/a$ and the width of the yellow strip becomes $d_1/a$ by assuming that we use exactly the same materials. We still assume that the shape of the thinner side of the strip is a circle segment. Since the equation is independent of the width of the strip, if we use the same materials with the same thickness, we obtain the same curvature $\kappa$ and the same radius $r$. 

Now, let $\theta'$ denote the new angle and it is given $d_1/a = r \theta'$. Since the original  $\theta = d_1/r$, the new angle can be computed in terms of the original angle as $\theta'=\theta/a$. Note that there are $a$ of them and each one of them has a width $c_{\mbox{thinner}} = 2 r sin \theta /2a $. Therefore, to keep both sides the same, we need to minimize 
$ c_{\mbox{longer}} - a c_{\mbox{thinner}} $. As a result, the following formula must be minimized if the result is negative. Small positive numbers are acceptable since this means an additional force is applied to keep the structure curved. By noticing that $r= d_1/theta$ we obtain: 
$$ d_1 \left( \frac{2}{\theta} \sin \frac{\theta}{2} - \frac{2a}{\theta} \sin  \frac{\theta} {2a} \right) $$

\paragraph{Remark 4:} Each term in this equation are in the form of $\sin \theta / \theta$, therefore, they both go to $1$ when $\theta$ goes to $0$. The function minimizes, but, this is not really good for our application, since when $\theta$ goes to $0$ $r$ goes to infinity and curvature becomes $0$, which corresponds to a straight line, which we do not want.  

\paragraph{Remark 5:} We want to point out that the formulas are in the form of $\sin 0.5 k \theta / (0.5 k \theta)$, where $k$ is any positive integer. In other words, we really deal with at most half of the original $\theta$. This makes the difference smaller. For instance,  For instance, for $\theta = 10 \mbox{degree}$, $0.5 \theta = 0.075 \mbox{radian}$, and $ \sin 0.5 \theta = 0.074929$. Let us assume that $a=10$ and $0.5\theta /a = 0.0075 \mbox{radian}$, then $ \sin \theta/a = 0.0074999$. The difference gives us 
$$ d_1 \left( \frac{0.074929}{0.075} - \frac{ 0.0074999}{0.0075} \right) = d_1 (0.99905333 - 0.99998666 ) \approx - 0.001 d_1$$
In other words, in this example, there will still be some force to make the structure less curved. Fortunately, this reactive force will be smaller than the bending force. 

\paragraph{Remark 6:} Increasing $a$ does not help since  the second term goes to $1$ by increasing $a$, and, as a result we get bigger difference. Therefore, it is better to reduce $a$. However, even reducing $a$ does not make much difference. For instance, in the previous example, let us change $a$ to $5$, then $0.5\theta /a = 0.015 \mbox{radian}$ and $ \sin \theta/a = 0.014999$. Now, the difference is computed as follows
$$ d_1 \left( \frac{0.074929}{0.075} - \frac{ 0.014999}{0.015} \right) = d_1 (0.99905333 - 9999625 ) \approx - 0.001 d_1$$ 
In other words, optimization will be a challenging problem since we deal with both continuous and discrete elements.

\section{Conclusion and Future Work}

In this paper, we have introduced a theoretical framework to design self-regulatory non-woven modules that can play the role of stomata in buildings. There is still a need for a significant amount of both theoretical and experimental work to turn this conceptual framework into practical self-regulating structures. We hope this paper will provide a strong interest in using non-woven structures in architecture and construction industries. 

This work only investigates the case of planar surfaces. The recent work on cyclic woven surfaces demonstrated that any surface can be covered with woven structures
\cite{akleman2009cyclic,akleman2011cyclic,xing2010single,akleman2020topologically,akleman2015extended}. It is straightforward to extend that work to non-woven cases. There already exists work to create woven tiles that can be used as stomata structures
\cite{yildiz2023modular,krishnamurthy2021geometrically}. 

This work was partially supported by the National Science Foundation under Grants NSF-CMMI-EAGER Award number 154824.

\bibliographystyle{unsrtnat}
\bibliography{references}
\end{document}